\documentclass[a4paper,10pt]{article}

\usepackage{epsf}
\usepackage{graphicx}    

\newcommand{\be}[1]{\begin{equation}\label{#1}}
\newcommand{\ee}{\end{equation}}
\newcommand{\bea}{\begin{eqnarray}}
\newcommand{\eea}{\end{eqnarray}}

\def\gsim{ \lower .75ex \hbox{$\sim$} \llap{\raise .27ex \hbox{$>$}} }
\def\lsim{ \lower .75ex \hbox{$\sim$} \llap{\raise .27ex \hbox{$<$}} }

\renewcommand{\markright}{\markright{\thepage}}

\begin{document}

\begin{titlepage}

\begin{flushright}
astro-ph/0609597
\end{flushright}

\vspace{5mm}

\begin{center}

{\Large \bf Observational $H(z)$ Data and Cosmological Models}

\vspace{10mm}

{\large Hao~Wei$^{\,1,}$\footnote{\,email address:
 haowei@mail.tsinghua.edu.cn} and Shuang~Nan~Zhang$^{\,1,2,3}$}

\vspace{5mm} {\em $^1$Department of Physics and Tsinghua Center for
 Astrophysics,\\ Tsinghua University, Beijing 100084, China\\
 $^2$Key Laboratory of Particle Astrophysics, Institute of High
 Energy Physics,\\ Chinese Academy of Sciences, Beijing 100049,
 China\\
 $^3$Physics Department, University of Alabama in Huntsville,
 Huntsville, AL 35899, USA}

\end{center}

\vspace{5mm}
\begin{abstract}
In this work, we confront ten cosmological models with
 observational $H(z)$ data. The possible interaction between dark
 energy and dust matter is allowed in some of these models. Also,
 we consider the possibility of (effective) equation-of-state
 parameter (EoS) crossing $-1$. We find that the best models have
 an oscillating feature for both $H(z)$ and EoS, with the EoS
 crossing $-1$ around redshift $z\sim 1.5$.\\

\end{abstract}

\end{titlepage}

\newpage

\setcounter{page}{2}

\section{Introduction}\label{sec1}

A lot of cosmological observations, such as SNe Ia~\cite{r1,r2},
 WMAP~\cite{r3}, SDSS~\cite{r4}, Chandra X-ray Observatory~\cite{r5}
 etc., find that our universe is experiencing an accelerated
 expansion. These results also suggest that our universe is spatially flat,
 and consists of about $70\%$ dark energy with negative pressure,
 $30\%$ dust matter (cold dark matters plus baryons), and negligible
 radiation. Dark energy study has been one of the most active fields in
 modern cosmology~\cite{r6}.

In the past years, many cosmological models are constructed to
 interpret the present accelerated expansion. One of the important
 tasks is to confront them with observational data. The most frequent
 method to constrain the model parameters is fitting them to the
 luminosity distance
 \be{eq1}
 d_L(z)=(1+z)\int_0^z \frac{d\tilde{z}}{H(\tilde{z})}\,,
 \ee
 which is an integral of Hubble parameter $H\equiv\dot{a}/a$, where
 $a=(1+z)^{-1}$ is the scale factor ($z$ is the redshift); a dot denotes
 the derivative with respect to cosmic time $t$. In this work, we
 use the observational $H(z)$ data directly, rather than
 the luminosity distance $d_L(z)$. The observational $H(z)$ data are
 based on differential ages of the oldest galaxies~\cite{r7}.
 In~\cite{r8}, Jimenez {\it et al.} obtained an independent estimate for
 the Hubble constant by the method developed in~\cite{r7}, and used it to
 constraint the equation-of-state parameter (EoS) of dark energy.
 The Hubble parameter depends on the differential age as a function
 of redshift $z$ in the form
 \be{eq2}
 H(z)=-\frac{1}{1+z}\frac{dz}{dt}\,.
 \ee
 Therefore, a determination of $dz/dt$ directly measures
 $H(z)$~\cite{r9}. By using the differential ages of passively
 evolving galaxies determined from the Gemini Deep Deep Survey
 (GDDS)~\cite{r10} and archival data~\cite{r11}, Simon {\it et al.}
 determined $H(z)$ in the range $0\,\lsim\, z\,\lsim\, 1.8$~\cite{r9}.
 The observational $H(z)$ data from~\cite{r9} are given in
 Table~\ref{tab1} and shown in Figs.~\ref{fig2}--\ref{fig5}.

\begin{table}[htbp]
\begin{center}
\begin{tabular}{c|lllllllll}\hline\hline
 $z$ &\ 0.09 & 0.17 & 0.27 & 0.40 & 0.88 & 1.30 & 1.43
 & 1.53 & 1.75\\ \hline
 $H(z)\ ({\rm km~s^{-1}\,Mpc^{-1})}$ &\ 69 & 83 & 70
 & 87 & 117 & 168 & 177 & 140 & 202\\ \hline
 $1 \sigma$ uncertainty &\ $\pm 12$ & $\pm 8.3$ & $\pm 14$
 & $\pm 17.4$ & $\pm 23.4$ & $\pm 13.4$ & $\pm 14.2$
 & $\pm 14$ &  $\pm 40.4$\\ \hline\hline
\end{tabular}
\end{center}
\caption{\label{tab1} The observational $H(z)$ data~\cite{r8,r9}
 (see~\cite{r15} also).}
\end{table}

These observational $H(z)$ data have been used to constrain the dark
 energy potential and its redshift dependence by Simon {\it et al.}
 in~\cite{r9}. Yi and Zhang used them to constrain the parameters of
 holographic dark energy model in~\cite{r12}. Some relevant works also
 include~\cite{r13,r14}. Recently, in~\cite{r15}, Samushia and Ratra
 have used these observational $H(z)$ data to constrain the parameters
 of $\Lambda$CDM, XCDM and $\phi$CDM models. In this work, we will
 revisit these observational $H(z)$ data and compare them with some
 cosmological models.

By looking carefully on the observational $H(z)$ data given in
 Table~\ref{tab1} and shown in Figs.~\ref{fig2}--\ref{fig5}, we
 notice that two data points near $z\sim 1.5$ and $0.3$ are very
 special. They deviate from the main trend and dip sharply, especially
 the one near $z\sim 1.5$; the $H(z)$ decreases and then increases
 around them. This hints that the effective EoS crossed $-1$ there.
 This possibility has not been discussed in previous
 works~(e.g.~\cite{r12,r15}). In the present work, we will seriously
 explore this possibility.


\begin{center}
\begin{figure}[htbp]
\centering
\includegraphics[width=0.45\textwidth]{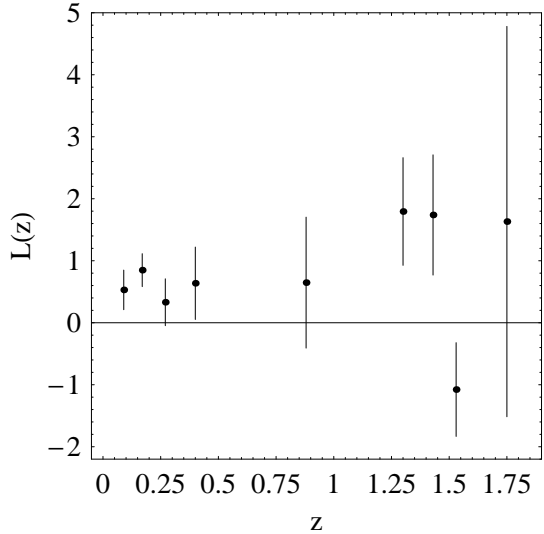}
\caption{\label{fig1} The quantity
 $L(z)\equiv H^2(z)/H_0^2-\Omega_{m0}(1+z)^3$ versus redshift $z$,
 for the fiducial parameters $H_0=72~{\rm km~s^{-1}\,Mpc^{-1}}$ and
 $\Omega_{m0}=0.3$.}
\end{figure}
\end{center}


On the other hand, we also consider the possible interaction
 between dark energy and dust matter. This is inspired by the data
 point near $z\sim 1.5$, which dips so sharply and stays clearly outside
 of the best-fit of the $\Lambda$CDM, XCDM and $\phi$CDM models studied
 in~\cite{r15}. In Fig.~\ref{fig1}, we show the quantity
 $L(z)\equiv H^2(z)/H_0^2-\Omega_{m0}(1+z)^3$ versus redshift $z$, which
 is associated with the fractional energy density of dark energy, for
 the fiducial parameters $H_0=72~{\rm km~s^{-1}\,Mpc^{-1}}$ and
 $\Omega_{m0}=0.3$, where the subscript ``0'' indicates the present
 value of the corresponding quantity. It is easy to see that the
 fractional energy density of dark energy of the point near
 $z\sim 1.5$ is negative (beyond $1\sigma$ significance). To
 avoid this, one can decrease the corresponding $\Omega_{m0}$ or
 make the matter decrease with the expansion of our universe slower
 than $a^{-3}$. Inspired by this, it is natural to consider the
 possibility of exchanging energy between dark energy and dust matter
 through interaction.

We assume that dark energy and dust matter exchange energy
 through interaction term $C$, namely
 \bea
 &&\dot{\rho}_{de}+3H\left(\rho_{de}+p_{de}\right)=-C,\label{eq3}\\
 &&\dot{\rho}_m+3H\rho_m=C,\label{eq4}
 \eea
 which preserves the total energy conservation equation
 $\dot{\rho}_{tot}+3H\left(\rho_{tot}+p_{tot}\right)=0$.
 The interaction forms extensively considered in the literature
 (see~\cite{r16,r17,r18,r19,r20,r21,r22,r23,r24,r25,r26,r39}
 for instance) are
 $$C\propto H\rho_m,\ H\rho_{tot},\ H\rho_{de},
 \ \kappa\rho_m\dot{\phi},\ \ldots$$
 In this work, we consider the simplest case for convenience, i.e.
 \be{eq5}
 C=3\alpha H\rho_m,
 \ee
 where $\alpha$ is a dimensionless constant. Combining
 Eqs.~(\ref{eq4}) and~(\ref{eq5}), it is easy to get
 \be{eq6}
 \rho_m=\rho_{m0}\,a^{-3(1-\alpha)}=\rho_{m0}(1+z)^{3(1-\alpha)}.
 \ee

In the following sections, we will compare the observational $H(z)$
 data with some cosmological models with/without interaction between
 dark energy and dust matter. We consider a spatially flat FRW
 universe throughout. We adopt the prior
 $H_0=72~{\rm km~s^{-1}\,Mpc^{-1}}$, which is exactly the median value
 of  the result from the Hubble Space Telescope (HST) key
 project~\cite{r27}, and is also well consistent with the one from
 WMAP 3-year result~\cite{r3}. Since there are only 9 observational
 $H(z)$ data points and their errors are fairly large, they cannot
 severely constrain model parameters alone. We perform a $\chi^2$
 analysis and compare the cosmological models to find out the one
 which catches the main features of the observational $H(z)$ data.
 We determine the best-fit values for the model parameters
 by minimizing
 \be{eq7}
 \chi^2(parameters)=\sum\limits_{i=1}^9\frac{\left[H_{mod}(parameters;z_i)
 -H_{obs}(z_i)\right]^2}{\sigma^2(z_i)}\,,
 \ee
 where $H_{mod}$ is the predicted value for the Hubble parameter in the
 assumed model, $H_{obs}$ is the observed value, $\sigma$ is the
 corresponding $1\sigma$ uncertainty, and the summation is over the
 9 observational $H(z)$ data points at redshift $z_i$.

\section{$\Lambda$CDM and XCDM models without/with interaction}\label{sec2}

In the spatially flat $\Lambda$CDM model, the Hubble parameter is
 given by
 \be{eq8}
 H(z)=H_0\sqrt{\Omega_{m0}(1+z)^3+(1-\Omega_{m0})}\,,
 \ee
 where $\Omega_{m0}\equiv\kappa^2\rho_{m0}/(3H_0^2)$ is the present
 fractional energy density of the dust matter, and
 $\kappa^2\equiv 8\pi G$. By minimizing the corresponding $\chi^2$,
 we find that the best-fit value for model parameter is
 $\Omega_{m0}=0.302$, while $\chi^2_{min}=9.04$ and
 $\chi^2_{min}/dof=1.13$, where $dof$ is the degree of freedom.

Then, we consider the interacting $\Lambda$CDM model
 (Int$\Lambda$CDM). In this case, Eq.~(\ref{eq3}) becomes
 \be{eq9}
 \dot{\rho}_\Lambda=-3\alpha H\rho_m.
 \ee
 By using Eq.~(\ref{eq6}), it is easy to find
 \be{eq10}
 \rho_\Lambda=\frac{\alpha}{1-\alpha}\,\rho_{m0}\,a^{-3(1-\alpha)}+const.
 \ee
 where $const.$ is an integral constant. Inserting into the
 Friedmann equation $H^2=\kappa^2(\rho_m+\rho_\Lambda)/3$ and
 requiring $H(z=0)=H_0$, one can determine the integral constant.
 Finally, in the Int$\Lambda$CDM model, the Hubble parameter is
 given by
 \be{eq11}
 H(z)=H_0\sqrt{\frac{\Omega_{m0}}{1-\alpha}(1+z)^{3(1-\alpha)}
 +\left(1-\frac{\Omega_{m0}}{1-\alpha}\right)}
 \ee
 By minimizing the corresponding $\chi^2$,
 we find that the best-fit values for model parameters are
 $\Omega_{m0}=0.386$ and $\alpha=0.138$, while $\chi^2_{min}=8.89$
 and $\chi^2_{min}/dof=1.27$.

\begin{center}
\begin{figure}[htbp]
\centering
\includegraphics[width=0.45\textwidth]{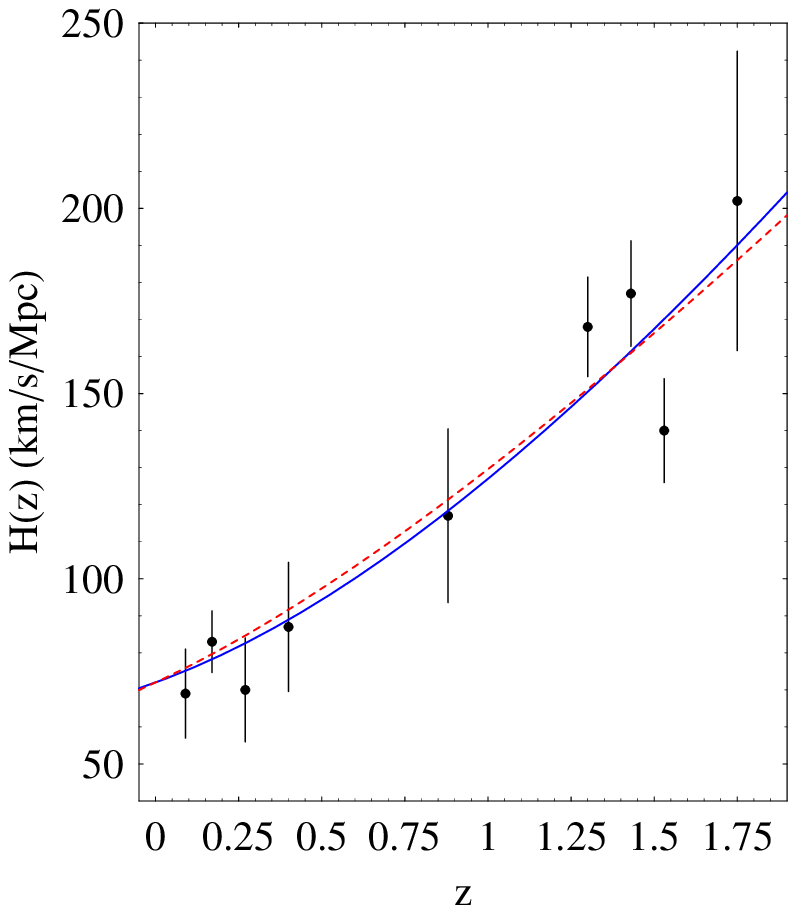}\hfill
\includegraphics[width=0.45\textwidth]{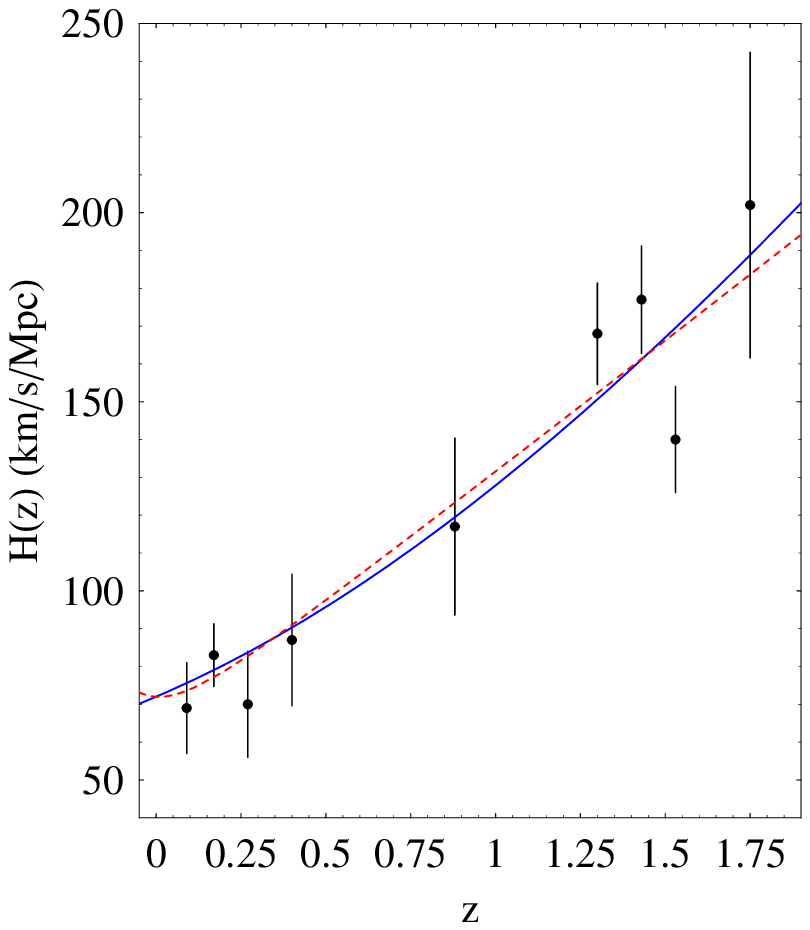}
\caption{\label{fig2} The observational $H(z)$ data with error bars,
 and the theoretical lines for $\Lambda$CDM (left panel) and XCDM (right panel)
 models with the corresponding best-fit parameters for the cases without
 (blue solid line) and with (red dashed line) interaction, respectively.}
\end{figure}
\end{center}

We present the observational $H(z)$ data with error bars, and the
 theoretical lines for $\Lambda$CDM (solid line) and
 Int$\Lambda$CDM (dashed line) models with the corresponding best-fit
 parameters in the left panel of Fig.~\ref{fig2}.

In the spatially flat XCDM model, the Hubble parameter is
 \be{eq12}
 H(z)=H_0\sqrt{\Omega_{m0}(1+z)^3+(1-\Omega_{m0})(1+z)^{3(1+w_X)}}\,,
 \ee
 where $w_X$ is the time-independent EoS of dark energy. By minimizing
 the corresponding $\chi^2$, we find that the best-fit values for model
 parameters are $\Omega_{m0}=0.284$ and $w_X=-0.899$, while
 $\chi^2_{min}=9.02$ and $\chi^2_{min}/dof=1.29$.


Next, we consider the interacting XCDM model (IntXCDM). In this
 case, Eq.~(\ref{eq3}) reads
 \be{eq13}
 \dot{\rho}_X+3H(1+w_X)\rho_X=-3\alpha H\rho_m.
 \ee
 Considering Eq.~(\ref{eq6}), we obtain
 \be{eq14}
 \rho_X=const.~a^{-3(1+w_X)}-\frac{\alpha\rho_{m0}}{\alpha+w_X}\,a^{-3(1-\alpha)},
 \ee
 where $const.$ is an integral constant. Again, inserting it into the
 Friedmann equation $H^2=\kappa^2(\rho_m+\rho_X)/3$ and
 requiring $H(z=0)=H_0$, we can determine this integral constant.
 Finally, in the IntXCDM model, the Hubble parameter is written as
 \be{eq15}
 H(z)=H_0\sqrt{\frac{w_X \Omega_{m0}}{\alpha+w_X}(1+z)^{3(1-\alpha)}
 +\left(1-\frac{w_X \Omega_{m0}}{\alpha+w_X}\right)(1+z)^{3(1+w_X)}}\,.
 \ee
 By minimizing the corresponding $\chi^2$, we find that the best-fit values
 for model parameters are $\Omega_{m0}=0.718$, $w_X=-3.705$ and
 $\alpha=0.302$, while $\chi^2_{min}=8.48$ and $\chi^2_{min}/dof=1.41$.

We present the observational $H(z)$ data with error bars, and the
 theoretical lines for XCDM (solid line) and IntXCDM (dashed line)
 models with the corresponding best-fit parameters in the right panel of
 Fig.~\ref{fig2}.

It can be seen clearly from Fig.~\ref{fig2} that none of the above
 four models may reproduce the sharp dip around $z\sim 1.5$; the
 data point near $z\sim 1.5$ deviates from model fitting by about
 $2\sigma$.

\section{Vector-like dark energy}\label{sec3}

As mentioned in the introduction, we are interested to seek a
 cosmological model whose effective EoS can cross $-1$. In this
 section, we consider the (interacting) vector-like dark energy
 model proposed in~\cite{r28} (see also~\cite{r16}). In this model,
 the EoS of vector-like dark energy and then the effective EoS
 can cross $-1$ in principle. From~\cite{r16,r28}, the energy density
 and pressure of vector-like dark energy are given by
 \bea
 &&\rho_A=\frac{3}{2}\left(\dot{A}+HA\right)^2+3V\left(A^2\right),\label{eq16}\\
 &&p_A=\frac{1}{2}\left(\dot{A}+HA\right)^2-3V\left(A^2\right)
 +2\frac{dV}{dA^2}A^2,\label{eq17}
 \eea
 where $A^2(t)$ is the time-dependent length of the so-called
 ``cosmic triad'' of three mutually orthogonal vector fields. In
 this work, we would like to consider the case with exponential
 potential, namely
 \be{eq18}
 V\left(A^2\right)=V_A\exp\left(-\lambda\kappa^2A^2\right),
 \ee
 where $V_A$ and $\lambda$ are constants. By using Eq.~(\ref{eq6})
 and new quantities $B\equiv\kappa A$ and
 $\tilde{V}_A\equiv\kappa^2V_A$, we can recast Eq.~(\ref{eq3}) as
 \be{eq19}
 \left(\dot{B}+HB\right)\left[\ddot{B}+3H\dot{B}+\left(\dot{H}+2H^2\right)B
 -2\lambda\tilde{V}_A B\,e^{-\lambda B^2}\right]
 =-3\alpha HH_0^2\Omega_{m0}(1+z)^{3(1-\alpha)}.
 \ee
 To find the solution for $B(t)$, we can use the initial conditions
 \be{eq20}
 B_{ini}=\pm\sqrt{\frac{2(\lambda-1)}{\lambda}},~~~~~~~
 \dot{B}_{ini}=0,
 \ee
 which come from~\cite{r16}, under the conditions $\lambda>1$ and
 $\alpha<1$. It is convenient to change the time $t$
 to redshift $z$, and
 \bea
 &&\dot{f}=-(1+z)Hf^\prime,\label{eq21}\\
 &&\ddot{f}=(1+z)H^2f^\prime-(1+z)\dot{H}f^\prime
 +(1+z)^2H^2f^{\prime\prime},\label{eq22}
 \eea
 where $f^\prime\equiv df/dz$ for any function $f$; we have used
 Eq.~(\ref{eq2}). From the Raychaudhuri equation
 $\dot{H}=-\kappa^2(\rho_A+\rho_m+p_A)/2$, we get the $\dot{H}$
 in Eq.~(\ref{eq22}) as
 \be{eq23}
 \dot{H}=-\left[-(1+z)B^\prime+B\right]^2H^2
 +\lambda\tilde{V}_A B^2 e^{-\lambda B^2}
 -\frac{3}{2}H_0^2\Omega_{m0}(1+z)^{3(1-\alpha)}.
 \ee
 From the Friedmann equation, the Hubble parameter reads
 $$H=H_0\sqrt{\Omega_{m0}(1+z)^{3(1-\alpha)}
 +\frac{\kappa^2\rho_A}{3H_0^2}}\,.$$
 Inserting Eq.~(\ref{eq16}) into it, we find
 \be{eq24}
 H^2=\frac{H_0^2\Omega_{m0}(1+z)^{3(1-\alpha)}+\tilde{V}_A
 e^{\lambda B^2}}{1-\frac{1}{2}\left[-(1+z)B^\prime+B\right]^2}.
 \ee
 From Eq.~(\ref{eq24}) and the requirement of $H(z=0)=H_0$, only
 three of $\alpha$, $\lambda$, $\Omega_{m0}$ and $\tilde{V}_A$ are
 independent of each other. We choose our free model parameters as
 $\alpha$, $\lambda$ and $\Omega_{m0}$.

Now, we can numerically solve Eqs.~(\ref{eq19}) and~(\ref{eq20})
 with the help of Eqs.~(\ref{eq21})--(\ref{eq24}) to get the $B(z)$.
 Then, we obtain the Hubble parameter $H(z)$ from Eq.~(\ref{eq24}).
 So, the corresponding $\chi^2$ is in hand.


\begin{center}
\begin{figure}[htbp]
\centering
\includegraphics[width=0.45\textwidth]{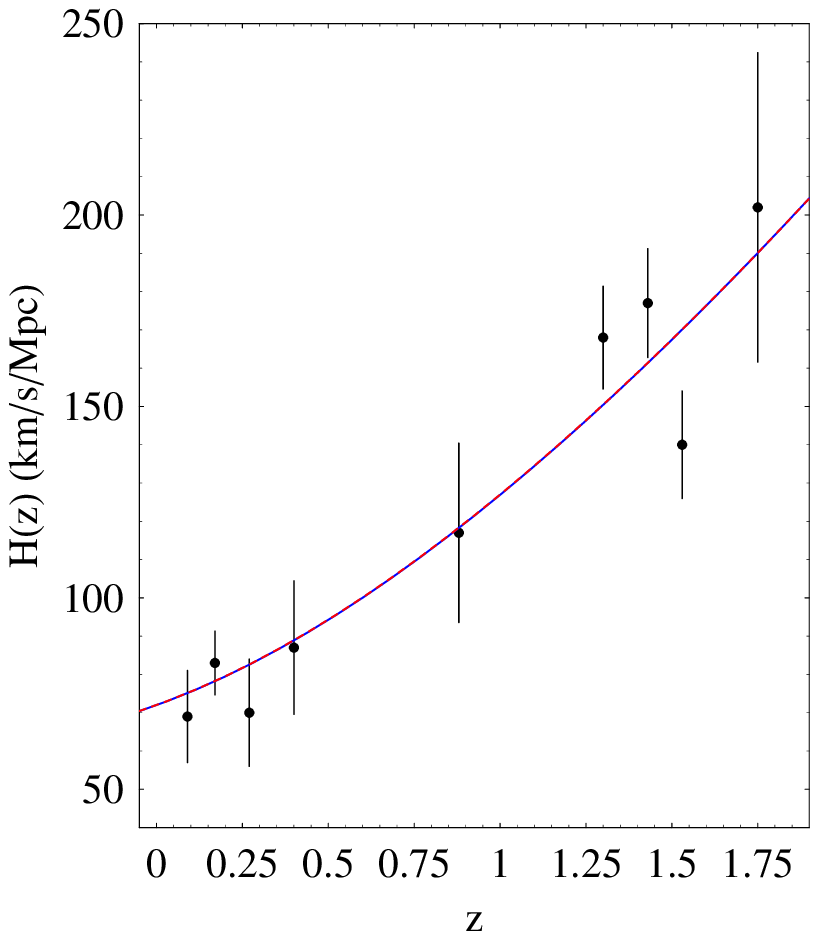}\hfill
\includegraphics[width=0.45\textwidth]{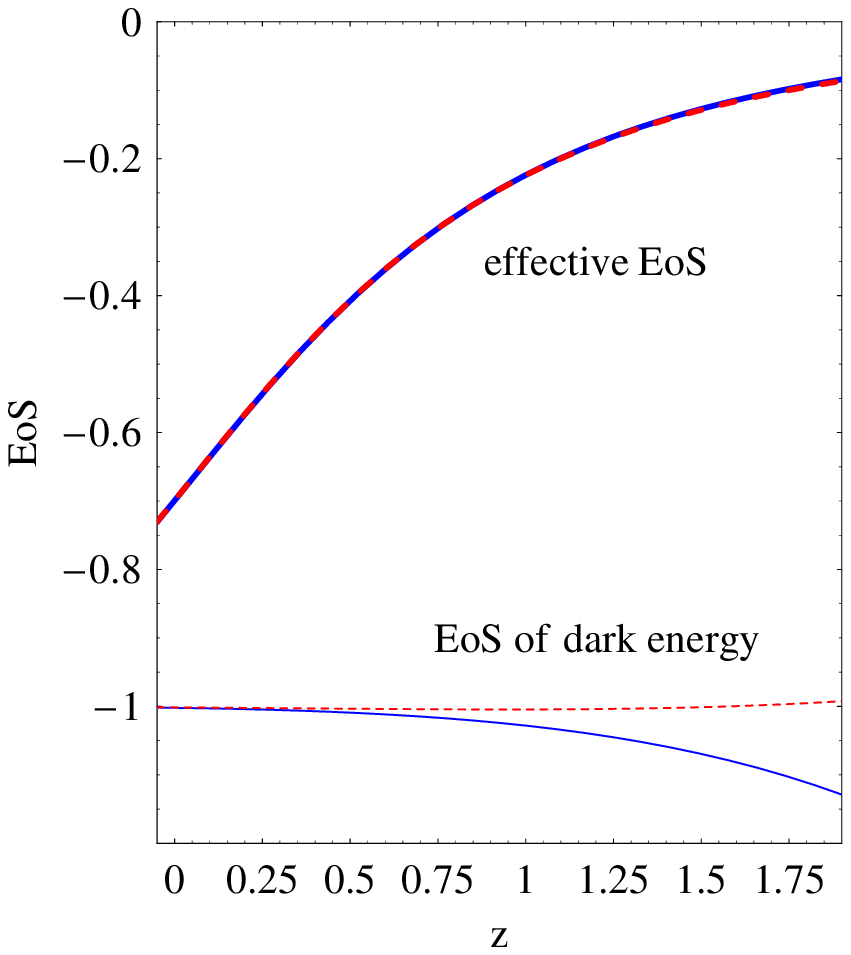}
\caption{\label{fig3} The left panel is the observational $H(z)$
 data with error bars, and the theoretical lines for VecDE (blue solid line)
 and IntVecDE (red dashed line) models with the corresponding best-fit
 parameters. The right panel shows the effective EoS (thick lines) and
 the EoS of vector-like dark energy (thin lines), for
 VecDE (blue solid line) and IntVecDE (red dashed line) models.}
\end{figure}
\end{center}


For the vector-like dark energy model without interaction (VecDE),
 i.e. $\alpha=0$ exactly, we find the minimal $\chi^2$ as
 $\chi_{min}^2=9.05$ ($\chi_{min}^2/dof=1.29$), for the best-fit
 parameters $\lambda=1.001$ and $\Omega_{m0}=0.299$ (the corresponding
 $\tilde{V}_A=0.70~H_0^2$).

For the interacting vector-like dark energy model (IntVecDE),
 i.e. leaving $\alpha$ as a free parameter, we find the minimal $\chi^2$
 as $\chi_{min}^2=9.04$ ($\chi_{min}^2/dof=1.51$), for the best-fit
 parameters $\alpha=-0.002$, $\lambda=1.005$ and $\Omega_{m0}=0.30$ (the
 corresponding $\tilde{V}_A=0.703~H_0^2$).

We present the observational $H(z)$ data with error bars, and the
 theoretical lines for VecDE and IntVecDE models with the corresponding
 best-fit parameters in the left panel of Fig.~\ref{fig3}. The right panel
 of Fig.~\ref{fig3} shows their corresponding effective EoS and
 the EoS of vector-like dark energy.

We see that the $H(z)$ predicted by VecDE and IntVecDE models, and
 their corresponding effective EoS, cannot be clearly distinguished.
 Also, it can be seen clearly from Fig.~\ref{fig3} that none of these two
 models may reproduce the sharp dip around $z\sim 1.5$; the data point
 near $z\sim 1.5$ deviates from model fitting by about $2\sigma$.
 Although the EoS of vector-like dark energy for VecDE and IntVecDE models
 are different, they have not crossed $-1$. We will briefly discuss this
 point in section~\ref{sec5}.

\section{The models with oscillating $H(z)$ ansatz}\label{sec4}

Obviously, all six models studied above fail to catch the features
 of the sharp dip around $z\sim 1.5$ and EoS crossing $-1$ mentioned in the
 introduction. In this section, we consider some parameterized models.

The first parameterized model (OA1) is the best model studied
 in~\cite{r29} which fits the SNe Ia data very well. Its Hubble
 parameter is given by
 \be{eq25}
 H(z)=H_0\sqrt{\Omega_{m0}(1+z)^3+a_1\cos (a_2 z^2+a_3)
 +\left(1-a_1\cos a_3-\Omega_{m0}\right)}\,,
 \ee
 where $a_1$, $a_2$ and $a_3$ are constants. By minimizing the
 corresponding $\chi^2$ for the $H(z)$ data, we find that the
 best-fit values for model parameters are $\Omega_{m0}=0.241$,
 $a_1=1.316$, $a_2=2.717$ and $a_3=-3.933$, while
 $\chi^2_{min}=4.27$ and $\chi^2_{min}/dof=0.85$.

Adding interaction into it, we obtain the IntOA1 model with Hubble
 parameter
 \be{eq26}
 H(z)=H_0\sqrt{\Omega_{m0}(1+z)^{3(1-\alpha)}+a_1\cos (a_2 z^2+a_3)
 +\left(1-a_1\cos a_3-\Omega_{m0}\right)}\,,
 \ee
 We find that the best-fit values for model parameters are
 $\Omega_{m0}=1.002$, $\alpha=0.407$, $a_1=-1.419$,
 $a_2=2.995$ and $a_3=-1.366$, while $\chi^2_{min}=3.54$ and
 $\chi^2_{min}/dof=0.89$.

\begin{center}
\begin{figure}[htbp]
\centering
\includegraphics[width=0.45\textwidth]{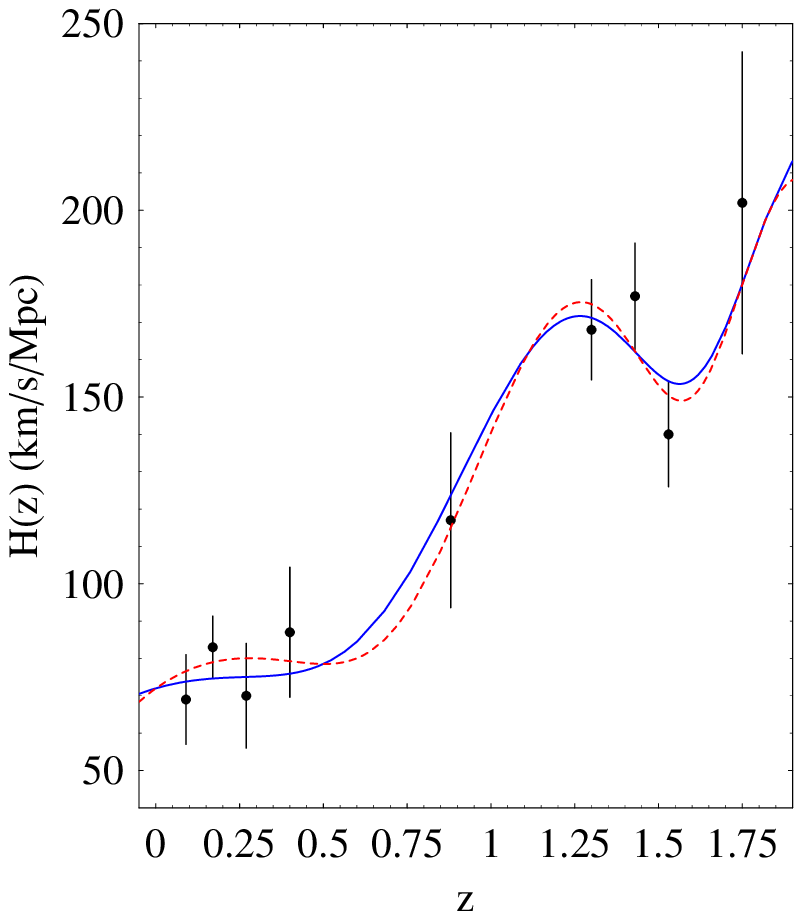}\hfill
\includegraphics[width=0.45\textwidth]{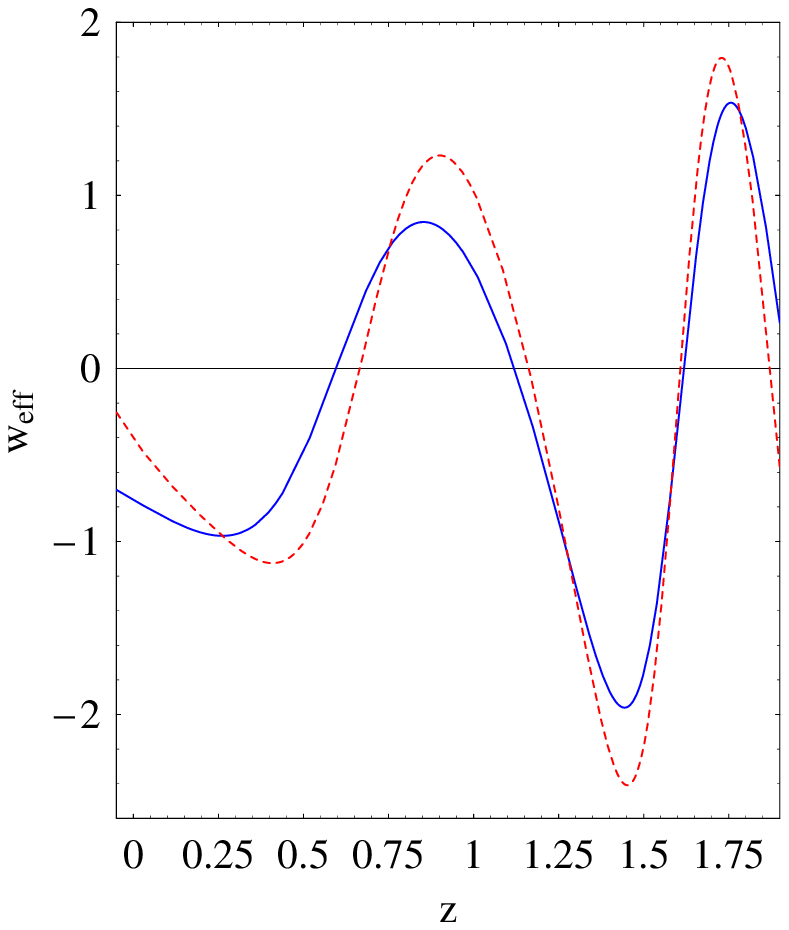}
\caption{\label{fig4} The left panel is the observational $H(z)$
 data with error bars, and the theoretical lines for
 OA1 (blue solid line) and IntOA1 (red dashed line) models with
 the corresponding best-fit parameters. The right panel shows
 their corresponding effective EoS.}
\end{figure}
\end{center}

We present the observational $H(z)$ data with error bars, and the
 theoretical lines for OA1 and IntOA1 models with the corresponding
 best-fit parameters in the left panel of Fig.~\ref{fig4}. The right
 panel of Fig.~\ref{fig4} shows the effective EoS,
 \be{eq27}
 w_{eff}\equiv\frac{p_{tot}}{\rho_{tot}}
 =-1+\frac{2}{3}(1+z)\frac{H^\prime}{H},
 \ee
 for OA1 and IntOA1 models, respectively. It is obvious that both
 effective EoS of OA1 and IntOA1 models crossed $-1$. The $H(z)$
 lines predicted by these two models dip around $z\sim 1.5$. So, it
 is not surprising that they fit the observational $H(z)$ data much
 better than the six models studied above.


Next, we consider a variant (OA2) of the best model studied
 in~\cite{r30} which also fits the SNe Ia data very well. Its Hubble
 parameter reads
 \be{eq28}
 H(z)=H_0\sqrt{\Omega_{m0}(1+z)^3+a_1(1+z)^3\left[\cos (a_2 z^2+a_3)
 -\cos a_3\right]+\left(1-\Omega_{m0}\right)}\,,
 \ee
 where $a_1$, $a_2$ and $a_3$ are constants. By minimizing the
 corresponding $\chi^2$, we find that the best-fit values for model
 parameters are $\Omega_{m0}=0.287$, $a_1=0.132$, $a_2=2.971$
 and $a_3=-4.481$, while $\chi^2_{min}=2.81$ and
 $\chi^2_{min}/dof=0.56$.

We get the IntOA2 model by adding interaction into OA2. The
 corresponding Hubble parameter is given by
 \be{eq29}
 H(z)=H_0\sqrt{\Omega_{m0}(1+z)^{3(1-\alpha)}
 +a_1(1+z)^{3(1-\alpha)}\left[\cos (a_2 z^2+a_3)
 -\cos a_3\right]+\left(1-\Omega_{m0}\right)}\,.
 \ee
 We find that the best-fit values for model parameters are
 $\Omega_{m0}=0.252$, $\alpha=-0.048$, $a_1=-0.117$,
 $a_2=2.975$ and $a_3=-1.351$, while $\chi^2_{min}=2.80$
 and $\chi^2_{min}/dof=0.70$.

\begin{center}
\begin{figure}[htbp]
\centering
\includegraphics[width=0.45\textwidth]{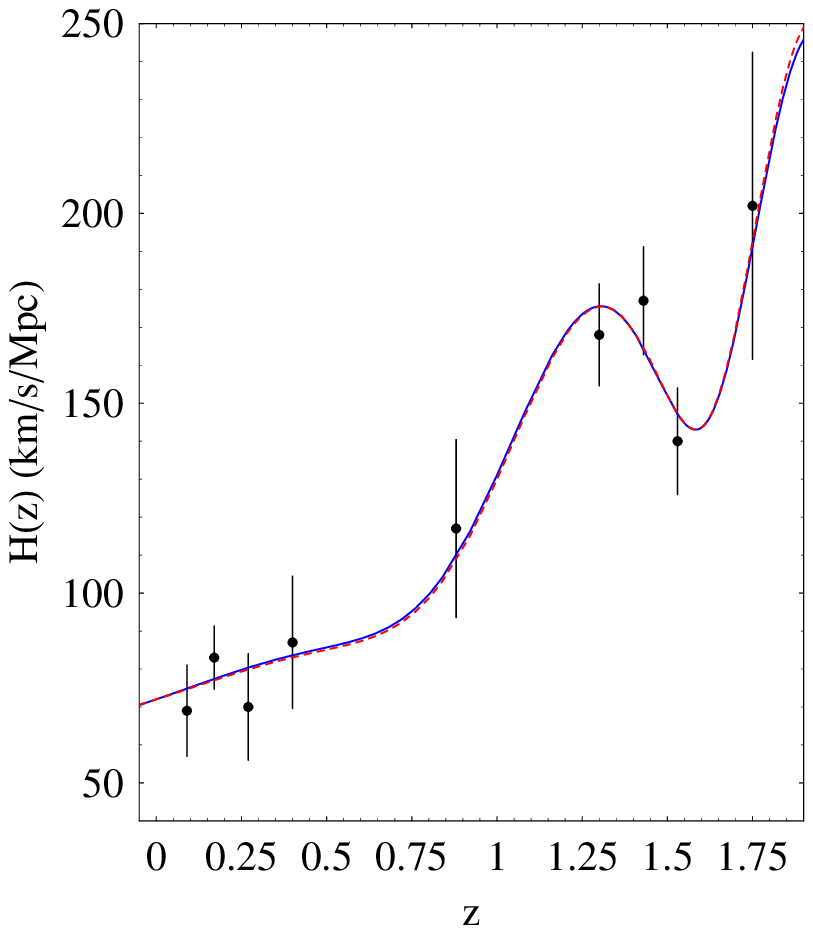}\hfill
\includegraphics[width=0.45\textwidth]{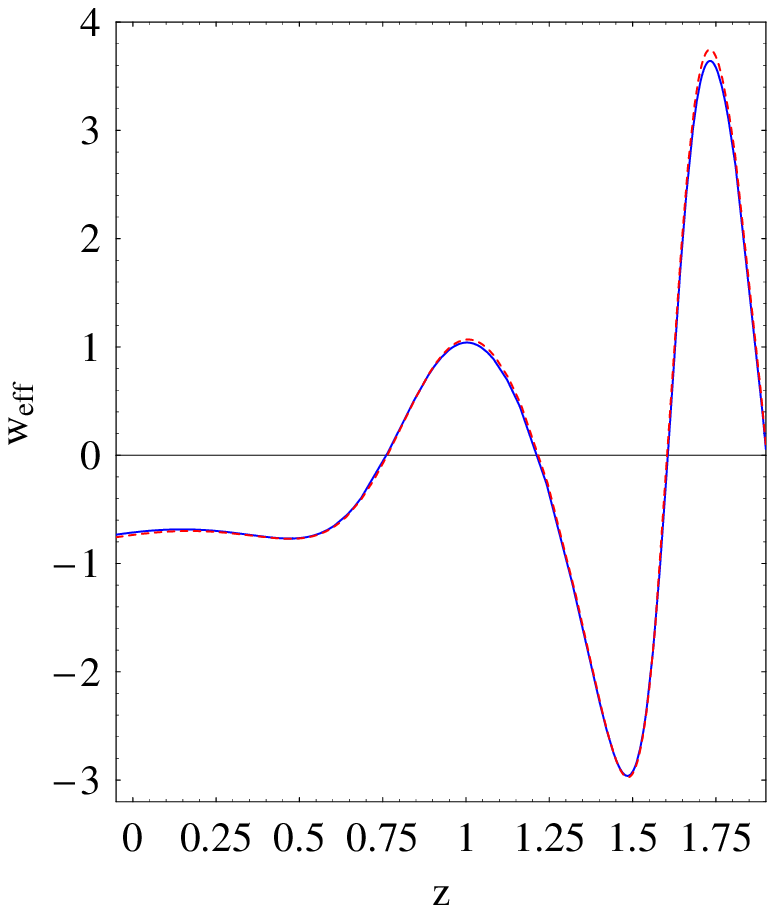}
\caption{\label{fig5} The left panel is the observational $H(z)$
 data with error bars, and the theoretical lines for
 OA2 (blue solid line) and IntOA2 (red dashed line) models with
 the corresponding best-fit parameters. The right panel shows
 their corresponding effective EoS.}
\end{figure}
\end{center}

We present the observational $H(z)$ data with error bars, and the
 theoretical lines for OA1 and IntOA1 models with the corresponding
 best-fit parameters in the left panel of Fig.~\ref{fig5}. The right
 panel of Fig.~\ref{fig5} shows their corresponding effective EoS.
 Again, we see the OA2 and IntOA2 models catch the main features of
 the observational $H(z)$ data; they reproduce the sharp dip around
 $z\sim 1.5$ and their effective EoS crossed $-1$ also. Since the
 value of $\alpha$ is too small, there is no significant difference
 between OA2 and IntOA2 models, unlike the case of OA1 and IntOA1 models.


\section{Concluding remarks}\label{sec5}

In Table~\ref{tab2}, we summarize all ten models considered in this
 work. It is obvious that they divide into two groups. The first six models
 fail to catch the main features of the observational $H(z)$ data,
 and hence have large $\chi_{min}^2$ and $\chi_{min}^2/dof$. The
 last four models with oscillating $H(z)$ ansatz, on the contrary, are
 well compatible with the observational $H(z)$ data. They can
 reproduce the sharp dip around $z\sim 1.5$, and their effective EoS
 crossed $-1$ also. Therefore, it is not surprising that their
 corresponding $\chi_{min}^2$ and $\chi_{min}^2/dof$ are significantly
 smaller than the ones of the first six models.

\begin{table}[htbp]
\begin{center}
\begin{tabular}{c|c|c|c|c}\hline\hline
 Model & $\hspace{4mm}\chi_{min}^2\hspace{4mm}$
 &$\ \chi_{min}^2/dof\ $
 & $P\left(\chi^2>\chi_{min}^2\right)$
 & $\begin{array}{c} {\rm Ranked\ by}\\
 P\left(\chi^2>\chi_{min}^2\right) \end{array}$\\ \hline
 $\Lambda$CDM & 9.04 & 1.13 & 0.34 & 5 \\
 Int$\Lambda$CDM & 8.89 & 1.27 & 0.26 & 6 \\
 XCDM & 9.02 & 1.29 & 0.25 & 7 \\
 IntXCDM & 8.48 & 1.41 & 0.21 & 9 \\
 VecDE & 9.05 & 1.29 & 0.25 & 8 \\
 IntVecDE & 9.04 & 1.51 & 0.17 & 10 \\ \hline
 OA1 & 4.27 & 0.85 & 0.51 & 3 \\
 IntOA1 & 3.54 & 0.89 & 0.47 & 4 \\
 OA2 & 2.81 & 0.56 & 0.73 & 1 \\
 IntOA2 & 2.80 & 0.70 & 0.59 & 2 \\
  \hline\hline
\end{tabular}
\end{center}
\caption{\label{tab2} Summarizing all ten models considered
 in this work.}
\end{table}

Some remarks are in order. The first one is on the
 failure of the vector-like dark energy model to reproduce the main
 features of the observational $H(z)$ data, since the EoS of vector-like
 dark energy and then the effective EoS can cross $-1$ in principle.
 According to~\cite{r16}, for the case of interaction term
 $C=3\alpha H\rho_m$, the vector-like dark energy behaves like a
 pure cosmological constant in the late time. So, it is not surprising
 that the vector-like dark energy model considered in this work is very
 close to the $\Lambda$CDM model. However, when the interaction term
 $C$ takes other forms, the late time behavior of vector-like dark
 energy can be considerablely different, as shown explicitly
 in~\cite{r16}. Therefore, it is interesting to compare the
 observational $H(z)$ data with the interacting vector-like dark energy
 model with other interaction forms and different potentials. We
 leave this to future works.

Secondly, we note that the best-fit value $\Omega_{m0}=1.002$
 of IntOA1 model is inconsistent with the results from clusters of
 galaxies~\cite{r38} and 3-year WMAP~\cite{r3} etc. In this sense, the
 IntOA1 model may be ruled out, although it fits the observational
 $H(z)$ data fairly well.

The third remark is concerned with EoS crossing $-1$. There are
 many pieces of other observational evidence for this in the
 literature~\cite{r30,r31,r32,r33}. Also, a lot of theoretical models
 whose EoS can cross $-1$ have been built (see for
 examples~\cite{r16,r17,r24,r32,r34,r40} and references therein). In
 this work, we present independent evidence for EoS crossed $-1$,
 from the observational $H(z)$ data.

Fourthly, we see that the last four models fit the data much better
 than the first six models. We note that the last four models have an
 oscillating feature, for both predicted $H(z)$ and EoS, as shown in
 Figs.~\ref{fig4} and~\ref{fig5}. In fact, this is also the main ideas
 of~\cite{r35,r36,r37,r41}. However, the last four models studied here
 and the ones considered in~\cite{r35,r36,r37,r41} are all
 {\em parameterized} models. We consider that it is important to
 build some physically motivated models in which the oscillating EoS
 arises naturally.

Fifthly, we admit that although the first six models have
 considerably larger $\chi^2_{min}/dof$, they are not unacceptable.
 For instance, the $\Lambda$CDM model has $\chi^2_{min}=9.04$ for
 eight degrees of freedom with $34\%$ probability, which is not
 unacceptably low. Also, before the new and improved $H(z)$ data are
 available, it is too early to talk about {\em strong observational
 evidence} for non-monotonic behavior of $H(z)$, since one can wonder
 that the dip around $z\sim 1.5$ could be due to unknown measurement
 errors and so on. Therefore, to firmly rule out non-oscillating
 models~(e.g. the first six models considered here), the more and
 better observational $H(z)$ data are required.

Finally, the observational $H(z)$ data provide an independent
 approach to constrain the cosmological models. However, by now, the
 observational $H(z)$ data only have 9 data points and their error bars
 are fairly large. Hence, they cannot severely constrain the cosmological
 models so far. A good news from~\cite{r15} is that a large amount of
 $H(z)$ data is expected to become available in the next few years.
 These include data from the AGN and Galaxy Survey (AGES) and
 the Atacama Cosmology Telescope (ACT), and by 2009 an order of
 magnitude increase in $H(z)$ data is anticipated.


\section*{Acknowledgments}
We thank Lado~Samushia and Raul~Jimenez for useful communications.
 We are grateful to Professor Rong-Gen~Cai for helpful discussions.
 We also thank Hui~Li, Xin~Zhang, Meng~Su, and Nan~Liang, Rong-Jia~Yang,
 Wei-Ke~Xiao, Jun-Zheng~Li, Yuan~Liu, Xiao~Che, Fu-Yan~Bian for kind help
 and discussions. We acknowledge partial funding support by the Ministry
 of Education of China, Directional Research Project of the Chinese Academy
 of Sciences and by the National Natural Science Foundation of China under
 project No.~10521001.


\end{document}